\documentclass[12pt]{JHEP3} % 10pt is ignored!
\usepackage{epsfig}
\usepackage{amssymb,amsfonts}
\usepackage{bbold}

\newcommand{\Tr}{{\rm Tr} }

\newcommand{\eins}{\mathbb{1}}            
\newcommand{\be}{\begin{equation}}
\newcommand{\ee}{\end{equation}}
\newcommand{\bea}{\begin{eqnarray}}
\newcommand{\eea}{\end{eqnarray}}
\newcommand{\ft}[2]{{\textstyle\frac{#1}{#2}}\,}
\newcommand\nn{\nonumber}
\newcommand{\eqn}[1]{(\ref{#1})}
\newcommand{\gdrei}{\gamma_{123}}

\newcommand{\sla}{a\!\!\!/\,}
\newcommand{\slb}{b\!\!/\,}
\newcommand{\Hint}{H_{\mbox{\tiny INT}}}
\def\xxx#1 {{\tt hep-th/#1}}

\newcommand{\rar}{\rightarrow}

\title{On the Spectrum of PP-Wave Matrix Theory}

\author{Nakwoo Kim and Jan Plefka\\
Max-Planck-Institut f\"ur Gravitationsphysik\\
Albert-Einstein-Institut\\
Am M\"uhlenberg 1, D-14476 Golm, Germany\\
{\tt kim,plefka@aei.mpg.de}}

\preprint{hep-th/0207034\\AEI-2002-047}  

\abstract{We study the spectrum of the recently proposed
matrix model of DLCQ M-theory in a parallel plane (pp)-wave 
background. In contrast to matrix theory in a flat background
this model contains mass terms, which lift the flat directions of
the potential and renders its spectrum discrete. 
The supersymmetry algebra of the model groups the energy eigenstates
into supermultiplets, whose members differ by fixed amounts of energy 
in great similarity to the representation of supersymmetry in
AdS spaces. There is a unique and exact zero-energy groundstate
along with a multitude of long and short multiplets of
excited states. 
For large masses the quantum mechanical model may be treated
perturbatively and we study the leading order energy shifts of 
the first excited states up to level two. Most interestingly
we uncover a {\it protected} short multiplet at level two, whose
energies do not receive perturbative corrections. Moreover, we
conjecture the existence of an infinite series of similar protected
multiplets in the pp-wave matrix model.}

\begin{document}

\section{Introduction}
The precise microscopic degrees of freedom of M-theory remain 
elusive even six years after its discovery \cite{M}. The most
promising candidate for such a description today is given by the
large $N$ limit of matrix theory \cite{bfss}, the maximally  
supersymmetric $U(N)$ gauge quantum mechanics \cite{QM,Rittenberg} 
which is intimately connected to the quantum supermembrane \cite{dWHN}.
The study of this seemingly simple model has been plagued, however, by
its nonlinearity and the existence of flat directions in the
potential leading to a continuous spectrum \cite{dWLN}.

Recently Berenstein, Maldacena and Nastase \cite{bmn} realized that
for eleven dimensional supergravity on the maximally 
supersymmetric parallel-plane (pp)-wave background \cite{ppwaves}
\bea
ds^2 &=& -2\, dx^+\, dx^- +\sum_{i=1}^9\, (dx^i)^2
-\Bigl ( \sum_{a=1}^3\, \ft{\mu^2}{9}\, (x^a)^2 +
 \sum_{a'=4}^9\, \ft{\mu^2}{36}\, (x^{a'})^2\, \Bigr )\, (dx^+)^2\nn\\
F_{123+}&=& \mu
\eea
the corresponding matrix theory is - opposed
to its $AdS_4\times S^7$ and $AdS_7\times S^4$ cousins 
\cite{backgroundfieldmem} - rather simple. In this model the flat
background  matrix model is augmented 
by bosonic and fermionic mass terms with a scale set by
$\mu$ along with a bosonic cubic interaction in 
the $SO(3)$ sector. As observed
in \cite{bmn} the mass terms remove the flat directions of 
the usual matrix theory potential and render 
its spectrum discrete. In fact the introduction of the mass
parameter $\mu$ into the model opens up a new perturbative window
of pp-wave matrix theory for $\mu\gg 1$ which we shall study in this
paper. First steps in this direction have been undertaken in
\cite{DSJvR}. Related work on the pp-wave matrix theory and
supermembrane may be found in \cite{pp}.

We begin our analysis after a careful statement of the model and
its quantization with the supersymmetry algebra, which due to 
the non-rigidness of the associated supersymmetry
variations displays some unusual features. 
The supersymmetry algebra groups the energy eigenstates
into multiplets, whose members do not have degenerate energy
eigenvalues any more, but differ by fixed amounts of energy
in great similarity to the representation of supersymmetry in
AdS spaces\footnote{For a recent review see e.g. \cite{Herger} and
references therein.}. 
The Hamiltonian naturally splits
into a free and an interacting piece in the limit $\mu\gg 1$, 
of which the free piece is given by a supersymmetric oscillator
system with vanishing groundstate energy protected from 
perturbative corrections. We then go on to study the leading 
order energy shifts of the first excited states in perturbation 
theory and find some surprises. In particular we uncover a
multiplet which does not receive any perturbative corrections
to its energy eigenvalues in leading order perturbation
theory. We argue that this result holds true to all orders.
Motivated by additional perturbative evidence, we are led
to conjecture the existence of an infinite series
of protected states in the full pp-wave matrix model.
Finally we end with some concluding remarks.

\section{The Model and its Quantization}
Our conventions are as follows: $X^i_{rs}$ and $\theta^\alpha_{rs}$ 
denote Hermitian $N\times N$ matrices, 
$i=1,\ldots , 9$ are the transverse vector indices
which split into $a=1,2,3$ and $a'=4,\ldots ,9$. Moreover for
the SO(9) Majorana spinors we work with a charge conjugation
matrix equaling unity, the Dirac matrices 
$\gamma^i_{\alpha\beta},\gamma^{ijkl}_{\alpha\beta}$ are symmetric
and $\gamma^{ij}_{\alpha\beta},\gamma^{ijk}_{\alpha\beta}$  are antisymmetric 
running over the spinor indices $\alpha,\beta=1,\ldots , 16$.
It is useful to perform a rescaling of $t\rightarrow \tau/(2R)$
of the time variable of the matrix model proposed in \cite{bmn} 
where $R$ denotes the radius of the compactified
direction in the DLCQ picture. With the help of this rescaling 
all parameters of  the matrix quantum mechanics are cast into the
single mass parameter $m=\mu/(2R)$. Then the pp-wave 
matrix model of \cite{bmn} takes the simple form
\be
S=S_{\mbox{\tiny flat}}+S_M
\label{lagrangian}
\ee
where
\bea
S_{\mbox{\tiny flat}}&=&\int d\tau\, \Tr \Bigl [ \ft 1 2 (DX^i)^2-i\theta D\theta
+\ft 1 4 [ X^i,X^j]^2+\theta \gamma_i\, [X^i,\theta]\, \Bigr ]
\nn\\
S_M&=& \int d\tau\, \Tr \Bigl [ -\ft 1 2 (\ft m 3 )^2\, (X^a)^2
-\ft 1 2 (\ft m 6 )^2\, (X^{a'})^2 +\ft m 4 i\, \theta\gdrei
\theta +\ft m 3 i \epsilon_{abc}\, X^a\,X^b\,X^c\, \Bigr ]\nn
\eea
and the covariant derivative is given by $D{\cal O}=\partial_\tau
{\cal O} -i [\omega,{\cal O}]$. 
It is invariant under the 16+16 linearly and non-linearly
realized supersymmetries
\bea
\delta X^i&=& 2\,\theta\gamma^i\epsilon(\tau) \nn\\
\delta \theta &=& \Bigl [ iDX^i\gamma_i+\ft 1 2 \, [X^i,X^j]\,
\gamma_{ij} +\ft m 3 i X^a\gamma_a\gdrei -\ft m 6 i X^{a'}\gamma_{a'}
\gdrei\, \Bigr ]\, \epsilon(\tau) + \eta(\tau)\nn\\
\delta \omega&=& 2\,\theta\epsilon(\tau) 
\label{SUSY}
\eea
with
\be \epsilon(\tau)= e^{-\frac{m}{12}\gdrei\, \tau}\, \epsilon_0
\qquad\qquad \eta(\tau)=e^{\frac{m}{4}\gdrei\, \tau}\, \eta_0 
\ee
Note the non-rigid character of the supersymmetry transformations:
The supersymmetry parameters depend explicitly on time.
This is the reason why the supercharge will be shown to not commute with
the Hamiltonian in the sequel. The cleanest way
to derive this model is to start from the supermembrane action
in $AdS_4\times S^7$ and $AdS_7\times S^4$ backgrounds given in 
\cite{backgroundfieldmem} and consider the pp-wave limit of
the superspace geometry along with the standard $\kappa$ 
gauge fixing condition ($\Gamma^+\theta=0$) for the fermions. 
The resulting membrane model may then be discretized in the
usual fashion \cite{dWHN} by approximating the group 
of area preserving diffeomorphisms by $U(N)$ in the limit 
$N\rightarrow \infty$. The outcome of this 
analysis is the model \eqn{lagrangian}. This derivation
is spelled out in detail in \cite{DSJvR}.
% and is
%more complete than the original derivation of
%\cite{bmn} where the pp-wave limit was taken only for 
%the superparticle action in an
%$AdS_7\times S^4$ background and the additional 
%cubic matrix model term was guessed.

It is straightforward to go to a Hamiltonian description of the
system. We choose the gauge $\omega=0$ and find that the resulting
Hamiltonian may be split into a free and an interacting piece
\be
H=H_0+\Hint
\label{Ham}
\ee
where
\bea
H_0&=& \Tr\Bigl [ \ft 1 2 (P^i)^2+ \ft 1 2 (\ft m 3)^2 \, (X^a)^2
+ \ft 1 2 (\ft m 6)^2 \, (X^{a'})^2 -\ft m 4 i \theta\gdrei\theta\, 
\Bigr ]\nn\\
\Hint &=& \Tr\Bigl [ -\ft m 3 i \epsilon_{abc}\, X^a\, X^b\,X^c
-\ft 1 4 [X^i,X^j]^2-\theta\gamma_i[X^i,\theta]\, \Bigr ]\nn
\eea
to be augmented by the gauge constraint
\be
{\cal G}=[P^i,X^i] -i\{\theta,\theta\} =0 \, .
\label{constraint}
\ee
As we shall show for $m\gg 1$ the interacting piece of the Hamiltonian $\Hint$
is suppressed and can be treated perturbatively.

Let us now turn to the quantization of the pp-wave matrix theory.
The canonical (anti)-commutation relations for the matrix operators
are given by
\be
[ P^i_{rs} , X^j_{tu} ] = -i\, \delta^{ij}\, \delta_{st}\, \delta_{ru}
\qquad
\{ \theta^\alpha_{rs} , \theta^\beta_{tu} \} = 
\ft{1}{2}\, \delta^{\alpha\beta}\, \delta_{st}\, \delta_{ru}
\ee
where the factor of $1/2$ for the fermions arises from the
Dirac procedure of treating the constraint $P^\alpha_\theta+i\theta^\alpha
=0$ properly. In view of $H_0$ in \eqn{Ham} it is natural to 
introduce the creation and
annihilation operators in the bosonic sector as
\be
a^a=\sqrt{\ft{3}{2m}}\, (P^a -i \ft m 3 X^a)
\qquad
b^{a'}=\sqrt{\ft{3}{m}}\, (P^{a'} -i \ft m 6 X^{a'})
\label{oscdef}
\ee
reflecting the $3+6$ split of the masses. They obey the standard
commutation relations
\be
[a^a_{rs},a^{\dagger\, b}_{tu}] =  \delta^{ab}\, \delta_{st}\, \delta_{ru}
\qquad
[b^{a'}_{rs},b^{\dagger\, b'}_{tu}] =  \delta^{a'b'}\, 
\delta_{st}\, \delta_{ru}\, .
\ee
Using these relations the bosonic part of the free Hamiltonian
$H_0$ takes the form
\be
H_0^B= \Tr\Bigl [ \, \ft m 3 \, a^{\dagger\, a}\, a^a +
\ft m 6 \, b^{\dagger\, a'}\, b^{a'}\, \Bigr ] + m\, N^2 \, .
\ee
In the fermionic sector we complexify the real spinor 
matrices $\theta_{rs}$ via
\be
\theta^\pm_{rs} = \Pi^\pm\, \theta_{rs} \quad \mbox{where}\quad
\Pi^\pm=\ft 1 2 \, (\eins \pm i\gdrei)
\label{complexify}
\ee
which yields the anticommutation relations
\be
\{ \theta^{+\,\alpha}_{rs} , \theta^{-\, \beta}_{tu} \} = 
\ft 1 2 \, (\Pi^+)_{\alpha\beta}\, \delta_{st}\, \delta_{ru}
\qquad \{ \theta^{+\,\alpha}_{rs} , \theta^{+\, \beta}_{tu} \} = 0=
\{ \theta^{-\,\alpha}_{rs} , \theta^{-\, \beta}_{tu} \} \, .
\ee
Note the chirality property of the complexified fermions
$(i\gdrei)\, \theta^\pm_{rs}=\pm \theta^\pm_{rs}$ as well as 
$(\Pi^\pm)^2=\Pi^\pm$ and $ \Pi^+\,\Pi^-=0$.
As $\theta=\theta^++\theta^-$ the fermionic term in $H_0$ is now
given by
\be
H_0^F=-i\ft m 4 \, \Tr\, [\, \theta\gdrei\theta\, ] = \ft m 2\, \Tr [\,  
\theta^{+\,\alpha}\, \theta^{-\,\alpha}\, ] -m\, N^2
\ee
canceling precisely the zero point energy of the bosonic sector.
The zero-energy groundstate of the resulting free Hamiltonian
\be
H_0 = \Tr\Bigl [ \, \ft m 3 \, a^{\dagger\, a}\, a^a +
\ft m 6 \, b^{\dagger\, a'}\, b^{a'} + \ft m 2\,
\theta^{+\,\alpha}\, \theta^{-\,\alpha}\,\Bigr ]
\ee
is denoted by $|0\rangle$ and is annihilated by
\be 
a^a_{rs}\, |0\rangle =0\qquad
b^{a'}_{rs}\, |0\rangle =0\qquad
\theta^{-\, \alpha}_{rs}\, |0\rangle =0\, .
\ee
Physical states are required to be gauge invariant due to the
gauge constraint \eqn{constraint}. They are given
by traces over words in the creation operators $a^\dagger, b^\dagger$
and $\theta^+$, viz.
\be
\Tr[\ldots a^{\dagger\, a}\ldots 
b^{\dagger\, a'}\ldots \theta^{+\, \alpha}\ldots]
\ldots
\Tr[\ldots a^{\dagger\, b}\ldots 
b^{\dagger\, b'}\ldots \theta^{+\, \beta}\ldots]
\, |0\rangle \,.
\ee
In this paper we shall be interested in the spectrum of the full
$U(N)$ Hamiltonian $H=H_0+ \Hint$. 
Clearly then the problem factorizes into the trivial free $U(1)$ 
sector spanned by the excitation operators of wordlength one,
$\Tr[a^{\dagger a}]$, $\Tr[b^{\dagger a'}]$, $\Tr[\theta^{+\, \alpha}]$,
and a complicated interacting $SU(N)$ sector spanned by excitation
operators of wordlength two and larger. 

\section{Supersymmetry Algebra and Structure of the Spectrum}

The derivation of the supersymmetry algebra is straightforward.
One has two supercharges $Q_\alpha$ and $q_\alpha$ associated with
the non-linearly and linearly realized supersymmetries of \eqn{SUSY}.
Their form follows from the operator relations
\bea
\delta X^i &=& 2i [ Q \,\epsilon (\tau) + q\,\eta(\tau), X^i ] 
\qquad \epsilon (\tau) = e^{-\frac{m}{12} \gamma_{123} \tau} \epsilon_0
\nonumber\\
\delta \theta_\alpha &=& 2i [ Q \,\epsilon (\tau) + q\,\eta(\tau)
, \theta_\alpha ] 
\,\qquad \eta (\tau) = e^{\frac{m}{4} \gamma_{123} \tau} \eta_0
\nonumber
\eea
One then deduces the two supercharges 
\bea
Q_\alpha &=& \Tr  \Bigl [\, [
P^i  \gamma_i - \ft i 2 [ X^i , X^j ]\, \gamma_{ij} 
+ \ft m 3 X^a\,  \gamma_a \gdrei 
+ \ft m 6 X^{a'} \gamma_{a'} \gdrei \,
]_{\,\alpha}{}^\gamma \, \theta_\gamma \, \Bigr ]\label{Q}\\
q_\alpha &=& \Tr \,[\, \theta_\alpha\, ]\, .
\label{q}
\eea
Note that $q_\alpha$ only acts in the free $U(1)$ sector of the
model. We relegate the explicit evaluation of the
supersymmetry algebra into appendix A. One finds in the
interacting  $SU(N)$ sector of the model
\bea
\{ Q_\alpha , Q_\beta \}
&=&  \delta_{\alpha\beta} H  
-\ft{m}{6} L^{ab} (\gamma_{ab}\,\gamma_{123})_{\alpha\beta} 
+\ft{m}{12} L^{a'b'} (\gamma_{a'b'}\,\gamma_{123})_{\alpha\beta} 
+ i \Tr ( X^i {\cal G} ) (\gamma_i)_{\alpha\beta} \nonumber \\
\relax [ H, Q_\alpha ] &=& \frac{m i}{12} (Q \gdrei)_{\alpha} + 
\Tr (\theta_\alpha {\cal G} ) 
\label{susalg}
\eea
Compared to the Minkowski background superalgebra we thus see 
the emergence of additional terms coupling to the angular momentum
operators $L^{ab}$ and $L^{a'b'}$ in the anti-commutator
of two supercharges, given by
\be
L^{ij} = \Tr ( \, P^i\, X^j - P^j\, X^i
 + \ft i 2 \theta \gamma^{ij} \theta )
\label{angmom}
\ee
and obeying
$
[L^{ij},L^{kl}] = -i\, ( \,\delta^{jk}\, L^{il} +\delta^{il}\, L^{jk}
-\delta^{ik}\, L^{jl} -\delta^{jl}\, L^{ik}\,  )$.
Despite the appearance of these angular momentum 
operators the old argument for the zero-energy groundstate still
goes through: a maximally supersymmetric state 
(being annihilated by all the $Q_\alpha$) will have zero energy
and be a $SO(3)$ and $SO(6)$ 
singlet. Hence the vanishing energy of the
groundstate $|0\rangle$ for $m\rar\infty$ is protected 
from perturbative corrections and constitutes the unique
groundstate of the interacting model. Moreover all excitations will have
strictly positive energy.

We see in \eqn{susalg} that
the supercharges do {\it not} commute with the Hamiltonian, which
simply states that superpartners do not have the same mass in this model. 
This effect is induced by the time dependent supersymmetry parameter,
and the coefficient $\ft{m}{12}$ can be easily inferred by computing
the difference of bosonic and fermionic masses $\ft m 3 -\ft m4 =\ft{m}{12}
=\ft m 4 - \ft m 6$. This phenomenon is
analogous to the situation for representations of
supersymmetry in AdS spaces.

The remaining (anti)-commutators read
\bea
\{ q_\alpha , q_\beta \}
&=& \ft N 2\, \delta_{\alpha\beta}\nn\\
\{Q_\alpha,q_\beta\} &=& \sqrt{\ft{2\,m}{3}}\, \Tr[(\sla\, \Pi^+)_{\alpha
\beta}+(\sla^\dagger\, \Pi^-)_{\alpha\beta}] +
\sqrt{\ft{m}{3}}\, \Tr[(\slb\, \Pi^+)_{\alpha
\beta}+(\slb^\dagger\, \Pi^-)_{\alpha\beta}]\nn\\
\relax  [H, q_\alpha ] &=& -\ft{m\, i}{4}\, (q\gdrei)_\alpha
\eea 
where the bosonic matrix ladder operators of \eqn{oscdef} appear in the
second line.

For the study of the spectrum of the pp-wave matrix model it is useful to perform
a chirality split of the dynamical supercharges $Q_\alpha$ according
to 
\be
Q^\pm = \Pi^\pm\, Q\qquad\mbox{where}\quad
 \Pi^\pm= \ft 1 2 ( \eins \pm i\gdrei) \, .
\ee
The supersymmetry algebra then takes the compact form
\bea
\{ Q^+_\alpha , Q^-_\beta \}
&=&  (\Pi^+)_{\alpha\beta} H  
+i\ft{m}{6} L^{ab} (\Pi^+\,\gamma_{ab})_{\alpha\beta} 
-i\ft{m}{12} L^{a'b'} (\Pi^+\, \gamma_{a'b'})_{\alpha\beta}  \nonumber \\
\{ Q^\pm_\alpha , Q^\pm_\beta \} &=&
0 \nonumber \\
\relax [ H, Q^\pm_\alpha ] &=& \mp \, \ft{m}{12} Q^\pm_{\alpha}
\label{susalg2}
\eea
where we have dropped the terms proportional to the gauge constraints
${\cal G}$ for transparency, as they do not affect gauge invariant
states.
It is instructive to spell out the form of $Q^\pm$ explicitly in terms
of oscillators
\bea
Q^-&=&\sqrt{\ft m 3}\, \Tr \, [\, b^{a'}\,\gamma_{a'}\, \theta^+]
+\sqrt{\ft {2m}{3}}\, \Tr \, [ \, a^{\dagger\,a}\,\gamma_{a}\, \theta^-]
-\ft i 2\, \Tr\, (\, [X^i,X^j]\, \gamma_{ij}\, \theta^-\, )\nn\\
Q^+&=&\sqrt{\ft {2m}{3}}\, \Tr \, [ \, a^{a}\,\gamma_{a}\, \theta^+]
+\sqrt{\ft m 3}\, \Tr \, [\, b^{\dagger\, a'}\,\gamma_{a'}\, \theta^-]
-\ft i 2\, \Tr\, (\, [X^i,X^j]\, \gamma_{ij}\, \theta^+\, )
\label{Qpm}
\eea
where $X^i$ is given in terms of the oscillators $a^{(\dagger)\,a}$ and 
$b^{(\dagger)\, a'}$
through \eqn{oscdef} and we have $(Q^+)^\dagger=Q^-$. Note
that the free theory supersymmetry algebra generated by $Q^\pm_0$
takes the same form as \eqn{susalg2}, with $H$ replaced by $H_0$ 
and $Q^\pm_0$ given by 
dropping the commutator terms in \eqn{Qpm}. Clearly now from \eqn{susalg2}
one observes that
$Q^+$ lowers and $Q^-$ raises the energy eigenvalue of a state
by $m/12$. As there are 8 raising and 8 lowering operators
a generic long multiplet will contain 256 states spread
over 9 ``floors'' of equal energies and 
spanning an energy range from its smallest value
$E$ to $E+\frac{2m}{3}$.

The simplest long multiplet is built upon a $SO(3)$ and $SO(6)$ singlet
on the "ground floor" and the entire multiplet has  256 states
in total. It is straightforward to find out how the states of such 
a multiplet are grouped into irreducible representations of $SO(3)\times SO(6)$
on each floor. The result reads
\be
\label{typical}
\begin{tabular}{cccccc}
Floor & \multicolumn{3}{c}{$SO(3)\times SO(6)$ reps} \cr
\hline
8
 &(1,1) & & \cr
7
 &$(2,\overline{4})$ & & \cr
6
 &$(1,\overline{10})$& $(3,\overline{6})$   \cr
5
 &$(2,\overline{20})$ & (4,4)& \cr
4
 &$(1,20')$ &(3,15) & (5,1)\cr
3
 &(2,20) &$ (4,\overline{4})$& \cr
2
& (1,10)& (3,6)  & \cr
1 & (2,4) && \cr
0 &(1,1) & & \cr
\end{tabular}
\ee
The energy differences within one multiplet are fixed, however
the lowest energy eigenvalue $E$ of a generic multiplet may
only be computed approximately in perturbation theory.

Let us now study the first excited states of the interacting
$SU(N)$ sector, i.e. excitations of wordlength two, which
are decomposed into irreducible representations as follows
\be
\begin{tabular}{lcc}
State & $SO(3)\times SO(6)$ rep & Energy\cr
\hline\\[-0.3cm]
$|a'a'\rangle = \Tr[b^{\dagger\, a'}\,b^{\dagger\, a'}\,]\, |0\rangle$
& $(1,1)$& $\ft m 3$ \cr
$|a'b'\rangle = 
%\Bigl( 
%\Tr[b^{\dagger\, a'}\,b^{\dagger\, b'}\,]
\Tr[b^{\dagger\, (a'}\,b^{\dagger\, b')}\,]
%-\delta^{a'b'}\frac{1}{6}\, \Tr[b^{\dagger\, c'}\,b^{\dagger\, c'}\,]
%\Bigr )
|0\rangle$
& $(1,20')$& $\ft m 3$ \cr
$|aa'\rangle_B=\Tr[a^{\dagger\, a}\,b^{\dagger\, b'}\,]\, |0\rangle$
&$(3,6)$&$\frac{m}{2}$\cr
$|aa'\rangle_F=\Tr[\theta^+\gamma^{aa'}\theta^+]\, |0\rangle$ & 
$(3,6)$& $\frac{m}{2}$\cr
$|a'b'c'\rangle=\Tr[\theta^+\gamma^{a'b'c'}\theta^+]\, |0\rangle$ & 
$(1,\overline{10})$& $\frac{m}{2}$\cr
$|aa\rangle = \Tr[a^{\dagger\, a}\,a^{\dagger\, a}\,]\, |0\rangle$
& $(1,1)$& $\ft {2m}{3}$ \cr
$|ab\rangle = 
%\Bigl( 
%\Tr[a^{\dagger\, a}\,a^{\dagger\, b}\,]
\Tr[a^{\dagger\, (a}\,a^{\dagger\, b)}\,]
%-\delta^{ab}\frac{1}{3}\, \Tr[a^{\dagger\, c}\,a^{\dagger\, c}\,]
%\Bigr )
|0\rangle$
& $(5,1)$& $\ft {2m}{ 3}$ \cr
 && \cr
$|a'a';\alpha\rangle = \Tr[\slb^{\dagger} \theta^+_\alpha\,]
|0\rangle$ & $(2,4)$ & $\ft {5m} {12}$\cr
$|a'b';\alpha\rangle = \Tr[\gamma^{(a'} b^{\dagger\, b')} \theta^+_\alpha\,]
|0\rangle$ & $(2,\overline{20})$ & $\ft {5m} {12}$\cr
$|aa;\alpha\rangle = \Tr[\sla^{\dagger} \theta^+_\alpha\,]
|0\rangle$  & $(2,\overline{4})$& $\ft {7m} {12}$\cr
$|ab;\alpha\rangle = \Tr[\gamma^{(a}a^{\dagger\, b)} \theta^+_\alpha\,]
|0\rangle$  & $(4,\overline{4})$& $\ft {7m} {12}$\cr
\end{tabular}
\label{wl2}
\ee
where the bifermion states are restricted to an odd number of $SO(6)$
vector indices, a consequence of the chirality property 
$\theta^+=\Pi^+\theta^+$. 
In our notation $(ij)$ refers to totally symmetrized indices without
the trace part.
Note that 
the states $|aa'\rangle_B$
and $|aa'\rangle_F$ are degenerate in mass and $SO(3)\times SO(6)$
representation and could potentially mix. Let us see how
these states fit into multiplets of the free superalgebra generated by
$Q_0^\pm$. As the free supercharges $Q_0^\pm$ preserve the 
wordlength it comes as no surprise, that the above states may
be grouped into two multiplets. Also due to
\bea
Q^+_0|a'a'\rangle=0&\qquad& Q^+_0|a'b'\rangle=0\nn\\
Q^-_0|aa\rangle=0&\qquad& Q^-_0|ab\rangle=0
\eea
the multiplets begin with the states of energy $m/3$ and
end with the states of energy $2m/3$ - they are short multiplets
consisting of 5 floors. The relevant double step 
ladder operators connecting floors of bosonic states are 
\bea
\mbox{2 floors up:}&\qquad  
Q^-_0\gamma^{aa'}Q^-_0\qquad Q^-_0\gamma^{a'b'c'}Q^-_0&\nn\\
\mbox{2 floors down:}&\qquad Q^+_0\gamma^{aa'}Q^+_0\qquad Q^+_0\gamma^{a'b'c'}Q^+_0&
\eea
which follow again from the chirality property of $Q^\pm_0$.
Starting with the lightest state of $|a'a'\rangle$  one finds
the multiplet ''{\bf A}'' 
\bea
|a'a'\rangle &\stackrel{Q^-_0\gamma^{aa'}Q^-_0}{\longrightarrow} 
(\, |aa'\rangle_F + \sqrt{2}\,|aa'\rangle_B\, )
\stackrel{Q^+_0\gamma^{aa'}Q^+_0}{\longleftarrow}& |ab\rangle\nn\\
|a'a'\rangle &\stackrel{Q^-_0\gamma^{a'b'c'}Q^-_0}{\longrightarrow} 
\qquad\qquad  0\qquad \qquad 
\stackrel{Q^+_0\gamma^{a'b'c'}Q^+_0}{\longleftarrow} &|ab\rangle
\eea
consisting of the 24 bosonic states $\{\,|a'a'\rangle,
(\, |aa'\rangle_F + \sqrt{2}\,|aa'\rangle_B\, ),|ab\rangle\, \}_{{\bf A}}$.
Similarly starting with $|a'b'\rangle$ as the lightest state 
the multiplet ``{\bf B}'' is obtained
\bea
|a'b'\rangle &\stackrel{Q^-_0\gamma^{aa'}Q^-_0}{\longrightarrow} 
(\, |aa'\rangle_F - 2\,\sqrt{2}\,|aa'\rangle_B\, )
\stackrel{Q^+_0\gamma^{aa'}Q^+_0}{\longleftarrow}& |aa\rangle\nn\\
|a'b'\rangle &\stackrel{Q^-_0\gamma^{a'b'c'}Q^-_0}{\longrightarrow} 
\quad\qquad|a'b'c'\rangle \quad \qquad
\stackrel{Q^+_0\gamma^{a'b'c'}Q^+_0}{\longleftarrow} &|aa\rangle
\eea
made out of the 49 bosonic states $\{\,|a'b'\rangle,
(\, |aa'\rangle_F -2\, \sqrt{2}\,|aa'\rangle_B\, ),|a'b'c'\rangle,
|aa\rangle\, \}_{{\bf B}}$. We indeed observe a mixing between the
two $(3,6)$ states $|aa'\rangle_B$ and $|aa'\rangle_F$, 
which is orthogonal due to the norms
\be
{}_F\langle a a'| b b' \rangle_F = 4\,\delta_{ab}\, \delta_{a'b'}\, N^2
\qquad
{}_B\langle a a'| b b' \rangle_B = \delta_{ab}\, \delta_{a'b'}\, N^2
\qquad
{}_F\langle a a'| b b' \rangle_B = 0 \, .
\label{norm1}
\ee
For the fermionic states it is obvious from  
$Q^-_0 |a'a'\rangle = |a'a';\alpha\rangle$ and 
$Q^+_0 |aa\rangle = |aa;\alpha\rangle$ that the fermionic
states $|a'a';\alpha\rangle$ and $|ab;\alpha\rangle$ belong
to the multiplet ``{\bf A}'', whereas
$|a'b';\alpha\rangle$ and $|aa;\alpha\rangle$ belong to ``{\bf B}''. 
So the level two states make
two irreducible supermultiplets, 
\bea
\mbox{{\bf A}:} &\quad& (1,1)+(2,4)+(3,6)+(4,\overline{4})+(5,1)\nn\\
\mbox{{\bf B}:} &\quad& (1,20')+(2,\overline{20})
+[(1,\overline{10})+(3,6)]+(2,\overline{4})+(1,1)\, .
\eea
We note that both of them can be part of the simplest long multiplet
presented in (\ref{typical}).  
In the next section we shall study how the energies of these multiplets
get corrected in perturbation theory.

%%%%%%%%%%%%%%%%%%%%%%%%%%%%%%%%%%%%%%%%%%%%%%%%%%%%%%%%%%
\section{The Perturbative Energy Spectrum and Protected States}
%%%%%%%%%%%%%%%%%%%%%%%%%%%%%%%%%%%%%%%%%%%%%%%%%%%%%%%%%%%a

The supersymmetry algebra derived in the last section implies
that the energy of the maximally supersymmetric ground state, 
which is annihilated by all supercharges, must be exactly zero.
Before we embark on the calculation of energy shifts for 
the excited states of \eqn{wl2} let us verify this in leading order
perturbation theory \footnote{This check was also performed in 
\cite{DSJvR}.}. The perturbative corrections to the spectrum are
organized in an expansion in $1/m^2$. To consistently work out the
leading correction of the groundstate energy 
it is then necessary to work up to second order
in quantum mechanical perturbation theory and to evaluate 
the expression
\be
\Delta E_0|_{{\cal O}(1/m^2)}= \langle 0|\,  \Hint|_{X^4}\, |0\rangle
+ \langle 0|\,  \Hint|_{X^3+X\theta^2}\, \frac{1}{E_0-H_0}\, 
\Hint|_{X^3+X\theta^2}\, |0\rangle
\label{series}
\ee
where the quartic interaction term contributes in first order
perturbation theory whereas the cubic and the Yukawa term 
contribute at second order perturbation theory.
The first term of the right-hand-side of \eqn{series} is given by the
expectation value
\be
\Delta E_0^1 = -\ft 1 4 \langle 0 | \, \Tr [X^i,X^j]^2 \, |0\rangle
= -\ft 1 2 \langle 0 | \, \Tr [X^i\,X^j\,X^i\,X^j-(X^i)^2\, (X^j)^2]\, |0\rangle
\ee
It is useful to maintain a unified language for the bosonic
ladder operators by introducing the objects
\be
X^i=-i\, (\, \tilde a^{\dagger\, i}- \tilde a^{i})
\label{uni1}
\ee
which obey
\be
[\tilde a^i_{rs}, \tilde a^{\dagger j}_{tu} ] = M^{ij}\,
\delta_{st}\, \delta_{ru}
\qquad
M^{ij}=\ft 3 m 
\,\left (\matrix{\ft 1 2\, \delta^{ab} & 0 \cr 0& \delta^{a'b'}}\right )
\label{uni2}
\ee
With the help of these definitions one straightforwardly
evaluates
\bea
\langle 0 | \, \Tr [X^i\,X^j\,X^i\,X^j ]\, |0\rangle&=&2N^3\, \Tr M^2 +N\, (\Tr
M)^2\nn\\
\langle 0 | \, \Tr[ (X^i)^2\, (X^j)^2]\, |0\rangle &=& N^3\, (\, \Tr M^2
+(\Tr M)^2\, ) + N\, \Tr M^2 
\eea
yielding the first contribution to the energy shift
\be
\Delta E_0^1 = -\ft 1 4 \langle 0 | \, \Tr [X^i,X^j]^2 \, |0\rangle
= \frac{11\cdot 3^4}{4\, m^2}\, N\, (N^2-1)
\label{eins}
\ee
The contributions to ${\cal O}(1/m^2)$ at second order perturbation
theory come from the cubic and the Yukawa terms of $\Hint$.
For the first contribution one has
\be
\Delta E_0^2 = \ft {m^2}{9}\,
\langle 0| \Tr [ e_{abc}\, \tilde a^a\, \tilde a^b\, \tilde a^c]\,
\frac{1}{m}\, \Tr [ e_{abc}\, \tilde a^{\dagger \,a}\, 
\tilde a^{\dagger\, b}\, \tilde a^{\dagger\, c}]\, |0\rangle
\ee
as here only a pure bosonic level 3 state is excited the free
Hamiltonian in the denominator has been replaced by $3\cdot\ft m 3$. 
Upon contracting this expression reduces to
\be
\Delta E_0^2 = - \frac{3^3}{4\, m^2}\, N(N^2-1)
\label{zwei}
\ee
Turning to the final contribution from the Yukawa coupling we have
\bea
\Delta E_0^3 &=& 
- \langle 0| \Tr (\theta^-\gamma_a\, [\tilde a^a,\theta^-] \,) \,
\frac{1}{\ft m 3 + \ft m 2 }\, \Tr (\, \theta^+\gamma_a\, [
\tilde a^{\dagger \,a},\theta^+\, ]\, ) \, |0\rangle\nn\\
&&- \langle 0| \Tr (\theta^-\gamma_{a'}\, [\tilde a^{a'},\theta^-] )\,
\frac{1}{\ft m 6 + \ft m 2 }\, \Tr (\,\theta^+\gamma_{a'} [
\tilde a^{\dagger \,a'},\theta^+\, ]\, ) \, |0\rangle
\label{Yuk}
\eea
Note the two different mass channels appearing for the inverse
free Hamiltonian in the above.
Now one computes
\be 
\langle 0| \Tr (\theta^-\gamma_i\, [\tilde a^i,\theta^-] \, )\,
 \Tr (\, \theta^+\gamma_i\, [
\tilde a^{\dagger \,i},\theta^+\, ]\, ) \, |0\rangle=
N\, (N^2-1)\, \mbox{tr} ( \gamma^i\Pi^-\gamma^j\Pi^+)\, M_{ij}
\ee
where $\Pi^\pm = \ft 1 2\, (\eins  \pm i\gdrei)$ are the projectors
appearing in \eqn{complexify}. We note that $\mbox{tr} 
( \gamma^a\Pi^-\gamma^a\Pi^+)=0$ and $\mbox{tr} 
( \gamma^{a'}\Pi^-\gamma^{a'}\Pi^+)=6\,\mbox{tr}\Pi^+$ which yields
the final result
\be
\Delta E_0^3 
=-\frac{8\cdot 3^3}{m^2} \,N\,(N^2-1)
\label{drei}
\ee
Summing up the three contributions \eqn{eins},\eqn{zwei} and
\eqn{drei} we indeed find
\be
\Delta E_0=\Delta E_0^1 +\Delta E_0^2 +\Delta E_0^3 =0
\ee
the vanishing shift of the groundstate energy in leading
order perturbation theory.

The computation of the energy shifts for the first excited
states of \eqn{wl2} goes along the same lines, but is
technically more involved. We report on the details of this
computation in appendix B and simply state the complete result
here. By virtue of the 
supersymmetry algebra \eqn{susalg2} it is clear that states within one
multiplet should receive the {\it same} perturbative correction
to their energy eigenvalues. This is indeed what one finds.
Taking care of the normalization of states the leading 
shift in energy for a generic state $|\phi\rangle$ is 
given by
\be
\Delta  E_{|\phi\rangle}\mid_{{\cal O}(1/m^2)} =
\frac{1}{\langle \phi|\phi\rangle}\, \Bigl (
\langle \phi|  \Hint|_{X^4}
+ \Hint|_{X^3+X\theta^2}\, \frac{1}{E_{0}-H_0}\, 
\Hint|_{X^3+X\theta^2}\, |\phi\rangle\, \Bigr )\, .
\label{diagstr}
\ee
For the states of our multiplet ``{\bf A}'' one obtains
$$
\Delta  E_{|\phi\rangle}\mid_{{\cal O}(1/m^2)} = 
\frac{108}{m^2}\, (N-\frac{1}{N})
$$
\be
\mbox{for}
\qquad
|\phi\rangle \in \Bigl\{ \,
h\, |a'a'\rangle,\,
h_{aa'}\,(\, |aa'\rangle_F + \sqrt{2}\,|aa'\rangle_B\, ),\,
h_{ab}\,|ab\rangle\,\Bigr \}
\label{shiftA}
\ee
introducing suitable polarization tensors $h,h_{aa'}$ and $h_{ab}$.
Most interestingly, however,  the energy shift for the members of our
multiplet ``{\bf B}'' precisely cancels!
$$
\Delta  E_{|\phi\rangle}\mid_{{\cal O}(1/m^2)} = 
0
$$
\be
\mbox{for}
\qquad
|\phi\rangle \in \Bigl\{ \,
h_{a'b'}\, |a'b'\rangle,\,
\tilde{h}_{aa'}\,(\, |aa'\rangle_F -2\, \sqrt{2}\,|aa'\rangle_B\, ),\,
g_{a'b'c'}\, |a'b'c'\rangle,\,
h\,|aa\rangle\,\Bigr \}
\label{shiftB}
\ee
We shall argue that this remains true to all orders in perturbation
theory. The crucial input from the representation theory of
Lie superalgebras here is that if a
multiplet is short its energy is quantized by the symmetry algebra. 
One simple way of seeing this
is to use the fact that the ground-floor state must be annihilated
by a product of less than nine supercharges, since short multiplets do not
span all of the 9 floors. One can calculate the norm of a state
of the form $(\prod_i Q^-_{\alpha_i}) |\Lambda\rangle$ using 
only the superalgebra
and obtain a set of linear equations involving $E$ and the Dynkin
labels of $SO(3)\times SO(6)$ of the ground-floor state $|\Lambda\rangle$. 
In fact the representation theory of Lie superalgebras is known
to some extent, original classifications and first important results 
are due to Kac \cite{kac}. A more detailed discussion of the 
representation theory of the M-theory
pp-wave superalgebra will be presented in a separate publication
\cite{jhp}.

Based on this insight, our calculation
implies that the multiplet ''{\bf A}'' should combine with other
short multiplets of the free theory to make a long multiplet
in the interacting theory for which there is no such
quantization rule from the symmetry algebra. The
multiplet ''{\bf B}'' on the other hand should stay short. 
Because there are no lighter states in the
free theory the ground-floor states $(1,1)$ and $(1,20')$ must
remain as the ground floor of the two multiplets also with interactions.
For the multiplet ''{\bf A}'' we see from (\ref{typical}) that
the first missing block is $(1,10)$ on the second floor 
with $E= \ft m 2$. In fact
this state is provided from the level 3 spectrum: 
$\Tr [b^{\dagger \,[a'} b^{\dagger \,b'} b^{\dagger \,c']}]|0\rangle$, which
also has $E=\ft m 2$. One can show that the free theory spectrum
can provide all the missing blocks of higher floors which are needed 
to complete ``{\bf A}'' into a long multiplet. 
In order to turn multiplet ''{\bf B}'' into a long multiplet 
starting from $(1,20')$ we would need the states
$(2,4)\times(1,20') = (2,\overline{20})+(2,60)$ on the first-floor, 
where in terms of Dynkin labels the $(60)$ of $SO(6)$ is 
given by $[1,2,0]$. But here, unlike the situation for the
multiplet ''{\bf A}'', these additional $(2,60)$ states
at $E_0=\ft {5m}{12}$ are simply not present in the free
spectrum, even if we consider states of higher level.
At $E_0=\ft {5m}{12}$ there are no other states than
the states listed in (\ref{typical}). So it turns out
that the multiplet ''{\bf B}'' is truly short even in the interacting
theory, and its energy is free from corrections to all orders.

The fact that the multiplet ``{\bf B}'' built upon the lightest state
$|a'b'\rangle$ is protected is strongly reminiscent of
the situation for the chiral primary
operators in ${\cal N}=4$ super Yang-Mills theory which
do not receive any radiative corrections to their scaling
dimensions. These operators are given by symmetric
traceless combinations of the six scalar fields 
$\Phi^I$ with $I=1,\ldots, 6$ of ${\cal N}=4$ super Yang-Mills theory, i.e.
\be
{\cal O}_n^A=C^A_{I_1 I_2\ldots I_n}\,
\Tr\, [\Phi^{I_1}\, \Phi^{I_2}\ldots \Phi^{I_n}]
\ee
 with $C^A_{I_1\ldots I_n}$
being totally symmetric in its lower indices and any contraction
among the lower indices vanishing, 
$C^A_{I_1\ldots I\ldots I\ldots I_n}=0$. Let us therefore consider
the multiplets built upon the lightest states
\be
|C_{(n)}\rangle = C_{a_1' a_2' \ldots a_n'}\, \Tr\, [b^{\dagger a_1'}\,
b^{\dagger a_2'}\ldots b^{\dagger a_n'}]\, |0\rangle
\label{Cn}
\ee
with $C_{a_1' a_2' \ldots a_n'}$ being totally symmetric and traceless,
i.e. $C_{a_1'\ldots a'\ldots a' \ldots a_n'}=0$. These states have
the mass $\frac{n\cdot m}{6}$ in the free theory. Clearly as
\be
Q_0^+|C_{(n)}\rangle=0
\ee
they constitute  the lightest state in a multiplet of the
free theory. It is tempting to speculate that the energy
eigenvalue of these states is protected from perturbative
corrections as well. We have computed the energy shifts 
for these states for $n=3,4$ and $5$ in leading order 
perturbation theory and indeed find that they cancel!
The explicit contributions are presented in appendix B.
Based on this evidence we therefore conjecture that 
all the states contained in the multiplets built on
\eqn{Cn} are protected and that their energy eigenvalues
are exactly given by the free theory values. The proof of
this conjecture should go along the same lines as the
arguments presented in the above for the case $|C_{(2)}\rangle$,
namely due to the absence of the required representations
in the free theory at higher mass levels.
We leave the detailed proof for future work.

%%%%%%%%%%%%%%%%%%%%%%%%%%%%%%%%%%%%%%%%%%%%%%%%%%%%%
\section{Conclusions and Outlook}
%%%%%%%%%%%%%%%%%%%%%%%%%%%%%%%%%%%%%%%%%%%%%%%%%%%%%

In this paper we have studied the spectrum of the 
recently found massive matrix quantum mechanics in a pp-wave background
and performed a second order perturbation calculation.
We uncovered a protected short multiplet of the theory whose
energy eigenvalues are now known exactly in the full interacting model.
Moreover we conjectured the existence of an infinite series of
such protected states.
Employing the matrix model conjecture
this is a non-trivial statement about the light-cone 
Hamiltonian of M-theory in a pp-wave background. 
In the case of the maximally supersymmetric
pp-wave solution of type IIB superstring, the precise energy spectrum
in the light cone gauge was presented by Metsaev \cite{metsaev}. 
Our results can be thought of as the M-theory counterpart. Using the
relation between the AdS space and the pp-wave, it was argued that
the string spectrum in the pp-wave background must be related to
the anomalous dimension of the dual CFT operator with large R-charge.
It is very tempting to conjecture the same correspondence between the
M-theory pp-wave solutions and the superconformal field theories 
of M2- and M5-branes. An intriguing fact is that the Penrose limit
of both $AdS_{4}\times S_{7}$ and $AdS_{7}\times S_{4}$ lead to 
the same pp-wave solution with $SO(3)\times SO(6)$ symmetry, implying
that they share essentially the same subsector. The M-brane field
theories are still largely mysterious but it would be very interesting
if we can compare the matrix theory calculation reported here with 
field theory calculations. 

There are a number of further interesting open question emerging. For example
the protected energy eigenvalues do not depend on $N$ 
and should therefore survive the large $N$
limiting procedure under which the matrix model
approximates the pp-wave supermembrane. What is the picture of
these states in the supermembrane theory? Furthermore, what can we
learn from these considerations for the notorious flat matrix model
in the limit $m\rightarrow 0$?

Finally, in our work we have exclusively studied the matrix model around the
``trivial'' vacuum $X^i=0$. As discussed in \cite{bmn,DSJvR}
there is a multitude of further maximally supersymmetric
vacua in the bosonic $SO(3)$ sector
corresponding to fuzzy sphere solutions of the equations of motion. 
As these vacua are
subject to the same superalgebra we expect that similar 
protected multiplets exist in these sectors of the theory as well.

\section*{Acknowledgments}
We would like to thank Gleb Arutyunov, Bernard de Wit, Kimyeong Lee, 
Jeong-Hyuck Park,
Soo-Jong Rey and Piljin Yi for useful discussions. NK thanks KIAS 
for hospitality during the ``KIAS Workshop on Strings and Branes''
where an earlier version of this paper was presented.

\begin{appendix}

\section{The Supersymmetry Algebra}

The nontrivial pieces of the supersymmetry algebra \eqn{susalg}
lie in the (anti)-commmutators involving $Q_\alpha$.
In computing the anticommutator of two supercharges 
$\{Q_\alpha, Q_\beta\}$ we shall only
focus on the terms proportional to $m$ as we know from the work of
 \cite{Rittenberg} how the $m$ independent terms work out.
One then has
\be
\{Q_\alpha, Q_\beta\} = [\ldots]_\alpha{}^\gamma{}_{rs}\,
 [\ldots]_\beta{}^\delta{}_{tu}\, 
\{\theta^{sr}_\gamma,\theta^{ut}_\delta\} -
\Bigl [\, [\ldots]_\alpha{}^\gamma{}_{rs} ,  [\ldots]_\beta{}^\delta{}_{tu}
\Bigr ]\, \theta^{ut}_\delta\, \theta^{sr}_\gamma
\label{QQ}
\ee
where we have used the abbreviation
$$
 [\ldots]_\alpha{}^\gamma{}_{rs}= [ P^i\, \gamma_i 
-\ft i 2 [X^i,X^j]\,\gamma_{ij}
+m_i\, X^i\, \gamma_i\, \gdrei\, \bigr ]_\alpha{}^\gamma{}_{rs}
$$
following from \eqn{Q}, $r,s,t,u$ denote U(N) matrix indices. 
In the above we have moreover employed an extended summation
convention for the index $i$ where
$$
m_i\, X^i\, \gamma_i\equiv \ft m 3\, X^a\, \gamma_a + \ft m 6\, X^{a'}\,
\gamma_{a'} 
$$ 
capturing the two different SO(3) and SO(6) masses in $m_i$.
The terms in \eqn{QQ} are then straightforwardly found to have the structure
\be
\{Q_\alpha, Q_\beta\} =\ft 1 2 (m_i\, X^i)^2 + 
+ {\cal B}_1 +{\cal B}_2
+ {\cal F} + (\mbox{$m_i$ independent terms})
\label{Struktur}
\ee
where
\bea
{\cal B}_1 &=& -\ft{1}{2}\,m_i\, \Bigl ( \Tr(P^k\, X^i)\, \gamma_k\gdrei
\gamma_i - \Tr(X^i\, P^k)\, \gamma_i\gdrei\gamma_k\Bigr )\nn\\
{\cal B}_2 &=& \ft i 4\, m_i\, \Tr(X^i\, [X^k,X^l]\, )\, 
\Bigr ( \gamma_i\gdrei\gamma_{kl}+ \gamma_{kl}\gdrei\gamma_i\, \Bigr )\nn\\
{\cal F} &=& i\,  \Bigl ( \gamma^i_{\alpha\gamma}\, 
(\gamma^i\gdrei)_{\beta\delta}\, m_i - (\gamma^i\gdrei)_{\alpha\gamma}\,
\gamma^i_{\beta\delta}\, m_i\, \Bigr )\, \Tr (\theta_\delta\, \theta_\gamma)
\eea
Now the bosonic contributions ${\cal B}_1$ and ${\cal B}_2$ can be shown
to be
\bea 
{\cal B}_1 &=&-\ft m 3\, \Tr(P^a\, X^b)\, (\gamma_{ab}\gdrei)_{\alpha\beta}
+\ft m 6 \, \Tr(P^{a'}\, X^{b'})\,  (\gamma_{a'b'}\gdrei)_{\alpha\beta}
\nn\\
{\cal B}_2 &=& -i\, \ft m 3\, \epsilon_{abc}\, \Tr(X^a\, X^b\,X^c)\,
\delta_{\alpha\beta}\nn
\eea
In order to reduce the fermionic contributions ${\cal F}$
in \eqn{Struktur} we make use of the Fierz identity
$$
\Tr(\theta_\delta\, \theta_\gamma) = 
\ft N 4\, \delta_{\delta\gamma} + \ft{1}{32}\, 
\Tr ( \theta\gamma^{jk}\theta)\, (\gamma_{jk})_{\delta\gamma}
+\ft{1}{96}\,\Tr ( \theta\gamma^{jkl}\theta)\, (\gamma_{jkl})_{\delta\gamma}
$$
which after some algebra gives us the relation
\be
{\cal F} =
-i\,\ft{m}{12} \, \Tr(\theta\gamma^{ab}\theta)\, 
(\gamma_{ab}\gdrei)_{\alpha\beta}
+i\,\ft{m}{24} \, \Tr(\theta\gamma^{a'b'}\theta)\, 
(\gamma_{a'b'}\gdrei)_{\alpha\beta}
-i \,\ft {m}{4}\, \Tr(\theta\gdrei\theta)\, \delta_{\alpha\beta}\,
\ee
We thus see the emergence of the angular momentum operators
\eqn{angmom}
in the algebra coupling to $(\gamma_{ab}\gdrei)_{\alpha\beta}$ and
$(\gamma_{a'b'}\gdrei)_{\alpha\beta}$ respectively.

Summarizing we then find the following anticommutator relation
\be
\{ Q_\alpha , Q_\beta \}
=  \delta_{\alpha\beta} H  
-\ft{m}{6} L^{ab} (\gamma_{ab}\,\gamma_{123})_{\alpha\beta} 
+\ft{m}{12} L^{a'b'} (\gamma_{a'b'}\,\gamma_{123})_{\alpha\beta} 
+ i \Tr ( X^i {\cal G} ) (\gamma_i)_{\alpha\beta}
\ee
putting back in the ``old'' $m$ independent terms computed in 
\cite{Rittenberg}. 

Now we want to calculate $[H, Q_\alpha]$. First
from the Jacobi identity one can easily show that
\be
[H , Q_{\alpha} ] = - \ft 1 8 [ \{ Q_\alpha, Q_\beta \}, Q_\beta ] 
\ee
Now using the above result for $\{Q_\alpha,Q_\beta\}$ and the property
\be
[ L^{ij} , S_\alpha ] = \ft i 2 S_\beta (\gamma^{ij} )_{\beta\alpha}
\ee
for any spinor, we obtain
\be
[H, Q_\alpha ] = \frac{m i}{12} (Q \gdrei)_{\alpha} + 
\Tr (\theta_\alpha {\cal G} )
\ee

\section{Details of the Perturbative Calculation}

In this section we comment on the calculation of energy shifts for 
the excited states. Naturally these manipulations are algebraically
more involved and we have performed them with the help of
{\tt Mathematica} and {\tt FORM} \cite{Jos} computer algebra systems.

The considered states of \eqn{wl2} which are excited by two raising
operators are conveniently expressed with the help of the
unified bosonic ladder operators of \eqn{uni1} and
\eqn{uni2} through
\bea
| h \rangle &=& h_{ij} \Tr 
(\tilde{a}^{\dagger \, i} \tilde{a}^{\dagger \, j} )
| 0 \rangle ,
\nonumber\\
| f \rangle &=& f_{aa'}
\Tr ({\theta^+} \gamma^{aa'} {\theta^+} )
| 0 \rangle ,
\nonumber\\
| g \rangle &=& g_{a'b'c'}
\Tr ({\theta^+} \gamma^{a'b'c'} {\theta^+} )
| 0 \rangle
\eea
The interactions of $\Hint$ respect the $SO(3)\times SO(6)$ split of the
free Hamiltonian, therefore mixing can only appear for the 
degenerate states $|aa'\rangle_B$ and $|aa'\rangle_F$. We will deal
with this problem at the end of this section and first study the
diagonal elements of the interactions in quantum
mechanical perturbation theory given by \eqn{diagstr}.

The states excited by bosonic oscillators will be dealt with first.
For the calculations done in this paper it turns out that the three
different interaction terms can be treated separately: the cross term
of the Yukawa and the cubic bosonic term does not contribute. So just like
the ground state we first consider the first order perturbation
of the quartic Yang-Mills interaction, and then the Yukawa and the
cubic bosonic term at second order. The results are summarized in a table
at the end of this section.

First the Yang-Mills quartic interaction is calculated to give
\bea
\label{E1}
\lefteqn{ -\frac{1}{4} \langle h_1| \Tr [ X^i , X^j ]^2 |h_2 \rangle = }
\nonumber \\
&& (N^5 - N^3) ( (\Tr M)^2 - \Tr M^2 ) \Tr h_1 M h_2 M
\nonumber\\
&& +4(N^3-N) ( \Tr h_1 M^2 \Tr h_2 M^2 - \Tr  h_1 M^2 h_2 M^2 )
\nonumber\\
&& +8(N^3-N) ( \Tr M \Tr h_1 M^2 h_2 M - \Tr h_1 M^3 h_2 M )
\eea
From this result one can easily check that there are no off-diagonal
overlaps of the pure bosonic states as expected.
Note that the normalization constants
for the different sets of states have  not been taken into account yet so
we have to divide the above result by the norm
\be
\langle h | h \rangle = 2N^2 \Tr (hMhM)
\ee
What we have to do now is simply calculate the ratios of 
traces of the form $\Tr (h M^p h M^q)$ which is straightforward.

Next we turn to the contributions from the cubic bosonic or Myers 
term. We write it in terms of raising and lowering 
operators and substitute
appropriate values for $E_0-H_0$. Written more explicitly
\bea
\Delta E^2 \langle h | h \rangle   &=&
\frac{m^2}{9}\,
\langle h| \Tr [ e_{abc}\, X^a\, X^b\, X^c]\,
\frac{1}{H_0-E_0}\, \Tr [ e_{abc}\, X^{a}\, 
X^{b}\, X^{c}]\, |h\rangle
\nonumber \\
&=& + \frac{m}{9}\,
\langle h| \Tr [ e_{abc}\, \tilde a^a\, \tilde a^b\, \tilde a^c]\,
\, \Tr [ e_{abc}\, \tilde a^{\dagger \,a}\, 
\tilde a^{\dagger\, b}\, \tilde a^{\dagger\, c}]\, |h\rangle
\nonumber \\
&& + 3m\,
\langle h| \Tr [ e_{abc}\, \tilde a^a\, \tilde a^b\, \tilde a^{\dagger\, c}]\,
\, \Tr [ e_{abc}\, \tilde a^{\dagger \,a}\, 
\tilde a^{\dagger\, b}\, \tilde a^{c}]\, |h\rangle
\nonumber \\
&& - 3m\,
\langle h| \Tr [ e_{abc}\, \tilde a^a\, 
\tilde a^{\dagger\, b}\, \tilde a^{\dagger\, c}]\,
\, \Tr [ e_{abc}\, \tilde a^{\dagger \,a}\, 
\tilde a^{b}\, \tilde a^{c}]\, |h\rangle
\nonumber \\
&& - \frac{m}{9}\,
\langle h| \Tr [ e_{abc}\, \tilde a^{\dagger\, ^a}\, 
\tilde a^{\dagger\, b}\, \tilde a^{\dagger\, c}]\,
\, \Tr [ e_{abc}\, \tilde a^{\dagger \,a}\, 
\tilde a^{b}\, \tilde a^{c}]\, |h\rangle
\eea
When we evaluate this it turns out that for level 2 states we are
interested in here only the first two channels have nonvanishing
contributions. The result is summarized as
\bea
\Delta E^2 &=&
-
\frac{ 36 {\rm Tr}_3 \,(MhMhM) + 12m^2
\epsilon^{abc} \epsilon^{def} M_{ad} 
(MhM)_{be} (MhM)_{cf} }{\Tr (hMhM) }
\left( \frac{N^2-1}{N m}  \right)
\nonumber \\
&& - \frac{27(N^3-N)}{4m^2} 
\eea
where ${\rm Tr}_3 X$ means one should take the trace of the 3 
dimensional part only, after calculating $X$ as a 9-dimensional matrix.
Again the result for different states can be found in table 1.

%\begin{center}
%\begin{table}[ht]
\begin{table}
\begin{center}
\label{spectrum}
%\caption{second order perturbation calculation of the energy spectrum}
%\begin{footnotesize}
\begin{tabular}{|l|c|c|c|c|c|}
\hline
\multicolumn{2}{|c|}{States} & Quartic & Myers & Yukawa & Total \cr
\hline
$|0 \rangle $&(1,1) & $\frac{891(N^3-N)}{4m^2}$ & $-\frac{27(N^3-N)}{4m^2}$ & 
$-\frac{864(N^3-N)}{4m^2}$ & 
0 \cr
\hline
$|aa\rangle$ & (1,1) & $\frac{891N^4-351N^2-540}{4m^2N}$& 
$-\frac{27(N^4+19N^2-20)}{4m^2N}$ & 
$-\frac{864(N^3-N)}{4m^2}$ & 
$0$ \cr
\hline
$|ab\rangle$ &(5,1) & $\frac{891N^4-405N^2-486}{4m^2N}$ & 
$-\frac{27(N^4+N^2-2)}{4m^2N}$ & 
$-\frac{864(N^3-N)}{4m^2}$ & 
$\frac{108(N^2-1)}{m^2N}$ \cr
\hline
$|a'a'\rangle$& (1,1) & $\frac{891N^4+405N^2-1296}{4m^2N}$& 
$-\frac{27(N^3-N)}{4m^2}$ & 
$-\frac{864(N^4-1)}{4m^2N}$ & 
$\frac{108(N^2-1)}{m^2N}$ \cr
\hline
$|a'b'\rangle$ &(1,20') & $\frac{891N^4-27N^2-864}{4m^2N}$& 
$-\frac{27(N^3-N)}{4m^2}$ & 
$-\frac{864(N^4-1)}{4m^2N}$ & 
$0$ \cr
\hline
$|aa'\rangle_B$ &(3,6) & $\frac{891N^3-207N-684}{4m^2N}$& 
$-\frac{27(N^4+3N^2-4)}{4m^2N}$ & $-\frac{432(2N^4-N^2-1)}{4m^2N}$ & 
$\frac{36(N^2-1)}{m^2N}$ \cr
\hline
$|aa'\rangle_F$ &(3,6) & $\frac{891(N^3-N)}{4m^2}$ & $-\frac{27(N^3-N)}{4m^2}$ & 
$-\frac{288(3N^4-4N^2+1)}{4m^2N}$ & 
$\frac{72(N^2-1)}{m^2N}$ \cr \hline
$|a'b'c'\rangle$ &(1,10) & 
$\frac{891(N^3-N)}{4m^2}$ & $-\frac{27(N^3-N)}{4m^2}$ & 
$-\frac{864(N^3-N)}{4m^2}$ & $0$ \cr
\hline
\end{tabular}
\caption{The diagonal contributions of the second 
order perturbation calculation of the energy spectrum according
to eq. (4.13). The states are defined in (3.10) and the
numbers in the parenthesis represent the associated
$SO(3)\times SO(6)$ representation. Only the states of represenation
$(3,6)$ receive off-diagonal contributions, which are evaluated in
(B.10).}
%\end{footnotesize}
%\end{table}
\end{center}
\end{table}

Now we can turn to the consideration of the Yukawa terms. For 
the states with bosonic oscillators only we can easily perform
the Wick contraction of the fermionic oscillators in the 
interaction term. The result contains
${\rm tr} ( \gamma^i \Pi^+ \gamma^j \Pi^- )M_{ij}$, and it is easy
to see that only the $SO(6)$ part of the Yukawa term gives nontrivial
contributions. So
\bea
\Delta E^3
\langle h | 
h \rangle &=& 
-\frac{3}{2m} 
\langle h | 
\Tr ( \theta^- \gamma_{a'} [ \tilde{a}^{a'} , \theta^- ] )
\Tr ( \theta^+ \gamma_{b'} [ \tilde{a}^{\dagger\, b'} , \theta^+ ] )
| h \rangle
\nonumber \\
&&
-\frac{3}{m} 
\langle h | 
\Tr ( \theta^+ \gamma_{a'} [ \tilde{a}^{\dagger\, a'} , \theta^+ ] )
\Tr ( \theta^- \gamma_{b'} [ \tilde{a}^{b'} , \theta^- ] )
| h \rangle
\nonumber \\
&=& 
-\frac{216}{m^2} (N^3-N) \langle h | h \rangle
-\frac{36}{m}
\langle h |
N \Tr ( \tilde{a}^{\dagger \, a'} \tilde{a}^{a'}) 
- \Tr \tilde{a}^{\dagger \, a'} \Tr \tilde{a}^{a'}
| h \rangle
\nonumber \\
&=& -\frac{216}{m^2} (N^3-N) \langle h | h \rangle
-\frac{144}{m}
(N^3-N) {\rm Tr}_6 \, (MhMhM)
\eea
where ${\rm Tr}_6$ means we take the trace of the six dimensional part only.

Now for the states excited by fermionic oscillators it is
clear that the contribution of the quartic and the Myers term 
must be the same as the ground state. The Yukawa interaction can
be considered as before, computing the different channels separately.
For a state $\Tr (\theta^+_\alpha \Gamma_{\alpha\beta} \theta^+_\beta) 
|0 \rangle$, we get
$$
\Delta E^{3} =
-\frac{54( 4 N^4 - 5N^2 + 1 )}{m^2N} 
+ \frac{18(N^2-1)}{m^2N} 
\frac{{\rm tr} ( \Gamma \gamma^a \Pi^+ \Gamma \gamma^a )}{{\rm tr} 
(\Gamma^2 \Pi^+)}
$$
For $\Gamma=\gamma^{aa'}$, 
${\rm tr} ( \Gamma \gamma^a \Pi^+ \Gamma \gamma^a )= 
{\rm tr} (\Gamma^2 \Pi^+)$, so
\be
\Delta E^3 = -\frac{72}{m^2} \left( 3 N^3 - 4 N + 1/N \right)
\ee
and for $\Gamma=\gamma^{a'b'c'}$ using 
${\rm tr} ( \Gamma \gamma^a \Pi^+ \Gamma \gamma^a )= 
-3{\rm tr} (\Gamma^2 \Pi^+)$, 
\be
\Delta E^3 = -\frac{216}{m^2} \left( N^3 -  N  \right)
\ee
Again this result is summarized in the table. 

What remains to be done
is the computation of the mixing of $|aa'\rangle_B$ and
$|aa'\rangle_F$ under perturbation theory. Here it is essential
to work with the properly normalized states
\be
{|aa'\rangle_{B_{\mbox{\tiny{norm}}}}}=\ft 1 N\, |aa'\rangle_B \qquad
{|aa'\rangle_{F_{\mbox{\tiny{norm}}}}}=\ft {1}{2\, N}\, |aa'\rangle_F 
\ee
following from \eqn{norm1}. For the cross term only the
Yukawa interaction piece contributes and one finds
\be
{}_{B_{\mbox{\tiny{norm}}}}\langle a a'| 
\Hint|_{X\theta^2}\, \frac{1}{E_{0}-H_0}\, 
\Hint|_{X\theta^2}  |b b'\rangle_{F_{\mbox{\tiny{norm}}}} 
= \sqrt{2}\, \frac{36\, (N^2-1)}{m^2\, N}
\, \delta_{ab}\,\delta_{a'b'} \, .
\ee
Combining this with the result quoted in table 1 one thus 
obtains the mixing matrix
\be
\frac{36\, (N^2-1)}{m^2\, N}\, 
\left ( \matrix{ 2 & \sqrt{2}\cr \sqrt{2} & 1\cr}
\right )
\ee
for the normalized states $|aa'\rangle_{F_{\mbox{\tiny{norm}}}}$ and
$|aa'\rangle_{B_{\mbox{\tiny{norm}}}}$.
It indeed has the eigenvalues $0$ and $\frac{108\, (N^2-1)}{m^2\, N}$
associated to the eigenvectors 
$(|aa'\rangle_{F_{\mbox{\tiny{norm}}}}
-\sqrt{2}|aa'\rangle_{B_{\mbox{\tiny{norm}}}})$ and
$(|aa'\rangle_{F_{\mbox{\tiny{norm}}}}
+\frac{1}{\sqrt{2}}|aa'\rangle_{B_{\mbox{\tiny{norm}}}})$
respectively. These are precisely the combinations appearing
in the $(3,6)$ sector of the free field multiplets ``{\bf A}'' and
``{\bf B}'' as stated in \eqn{shiftA} and \eqn{shiftB}.

Finally we turn to the totally symmetrized $SO(6)$ higher level states
$|C_{(n)}\rangle$ of \eqn{Cn}. The explicit contributions from the three sectors
of perturbation theory are stated in table 2 for $n=3,4,5$ which
add up to zero.

\begin{table}
\begin{center}
\footnotesize
\begin{tabular}{|l|c|c|c|c|}
\hline
State & Quartic & Myers & Yukawa & Sum \cr
\hline
$|C_{(3)}\rangle$ &  $\ft{297 N^6+ 432 N^4- 729 N^2}{4(m/3)^2}$
&$-\ft {9(N^6-N^2)}{4(m/3)^2}$
&$\ft{-288 N^6 -432 N^4 +720 N^2}{4(m/3)^2}$&
$0$ \cr
\hline
$|C_{(4)}\rangle$ &  $\ft{99 N^7+588 N^5-111 N^3-576 N}{(m/3)^2}$ &
$\ft{-3N^7-12N^5+15N^3}{(m/3)^2}$ &
$\ft{-96 N^7-576 N^5+96 N^3+576 N}{(m/3)^2}$ &
$0$ \cr
\hline
$|C_{(5)}\rangle$ & $\ft{15N^2(33N^6+542N^4+649N^2-1224)}
{4(m/3)^2}$ &
$\ft{-15 N^2(N^6+14N^4-7N^2-8)}{4(m/3)^2}$ &
 $\ft{-60 N^2 (2N^6+33N^4+41N^2-76)}{(m/3)^2}$ &
$0$ \cr
\hline
\end{tabular}
\caption{The vanishing of the second order perturbation calculation of the 
energy shifts of the  totally symmetrized $SO(6)$ higher level states
defined in eq. (4.17).}
\end{center}
\end{table}

During the calculation we have not distinguished connected
and disconnected diagrams, so for each type of interaction the
leading correction appears to be ${\cal O} ( \frac{N^3}{m^2} )$
for level-two states, however they always 
add up to zero due to the underlying supersymmetry.
We can thus see that the physical coupling constant in a large $N$
expansion is $\frac{N}{m^2}$.

\end{appendix}

%%%%%%%%%%%%%%%%%%%%%%%%%%%%%%%%%%%%%%%%%%%%%%%%%%%%%%%%%%%

\end{document}